\documentclass[a4paper,fleqn,usenatbib,useAMS]{mnras}
\usepackage[T1]{fontenc}
\usepackage[bottom]{footmisc}
\usepackage{ae,aecompl}
\usepackage{graphicx}	% Including figure files
\usepackage{amsmath}	% Advanced maths commands
\usepackage{amssymb}	% Extra maths symbols
\usepackage{multicol}        % Multi-column entries in tables
\usepackage{bm}		% Bold maths symbols, including upright Greek
\usepackage{pdflscape}	% Landscape pages
\usepackage[version=4]{mhchem}
\usepackage[T1]{fontenc}
\usepackage{xcolor}
\usepackage{adjustbox}
\usepackage{tikz}
\usepackage{rotating}
\usepackage{booktabs,caption}
\usepackage[flushleft]{threeparttable}
\usepackage{lscape}
\usepackage{graphicx,epsfig,epsf,graphics,epstopdf}
\usepackage{hyperref}
\usepackage{mhchem,hhline}
\usepackage[normalem]{ulem}
\usepackage{float}
\usepackage{amsmath,amstext,amssymb}
\usetikzlibrary{arrows}
\usetikzlibrary{positioning}
\usetikzlibrary{shapes,snakes}
\usepackage[T1]{fontenc}
\usepackage{ae,aecompl}
\usepackage{txfonts}

\title[Glycolaldehyde towards the G358.93--0.03 MM1]{Identification of the simplest sugar-like molecule glycolaldehyde towards the hot molecular core G358.93--0.03 MM1}

\author[Manna et al.]{Arijit Manna$^{1}$\thanks{E-mail:arijitmanna@mcconline.org.in}, Sabyasachi Pal$^{1}$\thanks{E-mail:sabya.pal@gmail.com}, Serena Viti$^{2,3}$, Sekhar Sinha$^{1}$\\
	$^{1}$Department of Physics and Astronomy, Midnapore City College, Paschim Medinipur, West Bengal, India 721129\\
	$^{2}$Leiden Observatory, Leiden University, PO Box 9513, 2300 RA Leiden, The Netherlands\\
	$^{3}$Department of Physics and Astronomy, UCL, Gower Street, London WC1E 6BT, UK
}

\begin{document}
	\label{firstpage}
	\pagerange{\pageref{firstpage}--\pageref{lastpage}}
	\maketitle
	
	% Abstract of the paper
	\begin{abstract}
Glycolaldehyde (\ce{CH2OHCHO}) is the simplest monosaccharide sugar in the interstellar medium, and it is directly involved in the origin of life via the ``RNA world'' hypothesis. We present the first detection of glycolaldehyde (\ce{CH2OHCHO}) towards the hot molecular core G358.93--0.03 MM1 using the Atacama Large Millimeter/Submillimeter Array (ALMA). The calculated column density of \ce{CH2OHCHO} towards G358.93--0.03 MM1 is (1.52$\pm$0.9)$\times$10$^{16}$ cm$^{-2}$ with an excitation temperature of 300$\pm$68.5 K. The derived fractional abundance of \ce{CH2OHCHO} with respect to \ce{H2} is (4.90$\pm$2.92)$\times$10$^{-9}$, which is consistent with that estimated by existing two-phase warm-up chemical models. We discuss the possible formation pathways of \ce{CH2OHCHO} within the context of hot molecular cores and hot corinos and find that \ce{CH2OHCHO} is likely formed via the reactions of radical HCO and radical \ce{CH2OH} on the grain surface of G358.93--0.03 MM1.
	\end{abstract}
	
	
	%%%%%%%%%%%%%%%%%%%%%%%%%%%%%%%%%%%%%%%%%%%%%%%%%%
	
	%%%%%%%%%%%%%%%%% BODY OF PAPER %%%%%%%%%%%%%%%%%%
	
	% The MNRAS class isn't designed to include a table of contents, but for this document one is useful.
	% I, therefore, have to do some kludging to make it work without masses of blank space.
	\begingroup
	\let\clearpage\relax
	%\tableofcontents
	\endgroup
	\newpage
	
\section{Introduction}
\label{sec:intro}
In the interstellar medium (ISM), glycolaldehyde (\ce{CH2OHCHO}) is known as one of the simplest aldehyde sugar, and it is the only sugar detected in space \citep{hal04}. The monosaccharide sugar molecule \ce{CH2OHCHO} is an isomer of both methyl formate (\ce{CH3OCHO}) and acetic acid (\ce{CH3COOH}) \citep{bel09, min20}. \ce{CH2OHCHO} is one of the important interstellar organic molecule in the ISM because when \ce{CH2OHCHO} reacts with propenal (\ce{CH2CHCHO}), it forms ribose (\ce{C5H10O5}) \citep{bel09}. Ribose (\ce{C5H10O5}) is known as the central constituent of RNA, and it is directly involved in the hypothesis of the origin of life in the universe. The organic molecule \ce{CH2OHCHO} also has a major role in the formation of three, four, and five-carbon sugars \citep{hal06}. The molecular lines of \ce{CH2OHCHO} were first detected towards Sgr B2 (N) \citep{hal00, hal01, hal04, hal06, bel13, xu19} and subsequently towards the hot molecular core G31.41+0.31 \citep{bel09, cal14, riv17, min20}, the solar-like protostar IRAS 16293--2422 B \citep{jo12, riv19}, the class 0 protostar NGC 7129 FIRS 2 \citep{flu14}, and the hot corino NGC 1333 IRAS2A \citep{cou15}. Recently,  emission lines of \ce{CH2OHCHO} were also tentatively detected from the hot molecular core G10.47+0.03 \citep{mondal21}.

 \begin{figure*}
 	\centering
 	\includegraphics[width=0.8\textwidth]{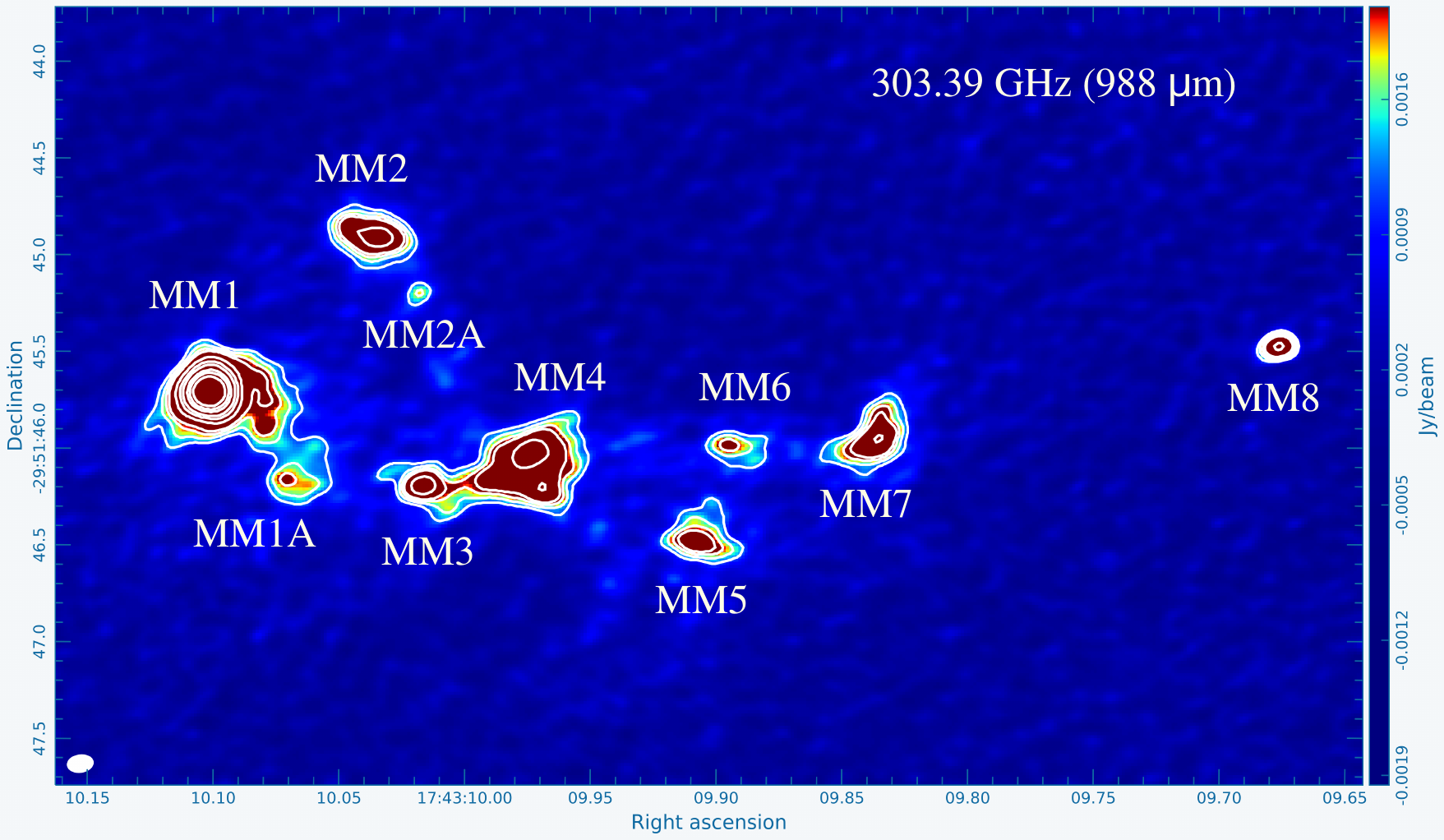}
 	\caption{Continuum emission image of high-mass star-formation region G358.93--0.03 at a wavelength of 988 $\mu$m. The synthesized beam size (white circle) of the continuum image is 0.412$^{\prime\prime}$$\times$0.363$^{\prime\prime}$. The sub-millimeter continuum sources MM1--MM8 are located in the massive star-formation region G358.93--0.03. The contour (white colour) levels start at 3$\sigma$ where $\sigma$ is the RMS of the continuum emission image and the contour levels increase by a factor of $\surd$2.}
 	\label{fig:cont}
\end{figure*}

The chemistry of hot cores is characterized by the sublimation of ice mantles, which accumulate during the star formation process \citep{shi21}. In prestellar cores and cold molecular clouds, the atoms and gaseous molecules freeze onto the dust grains. During the process of star-formation, the thermal energy and pressure increase due to the gravitational collapse. Therefore, dust temperatures also increase, and chemical interactions between heavy species become active on the grain surfaces \citep{gar06, shi21}. This leads to the formation of large complex organic molecules \citep{gar06, shi21}. In addition, sublimated molecules such as ammonia (\ce{NH3}) and methanol (\ce{CH3OH}) are also subject to further gas-phase reactions \citep{nom04, taq16, shi21}. Consequently, the warm and dense gas surrounding the protostars becomes chemically rich, resulting in the formation of one of the strongest and most powerful molecular line emitters known as hot molecular cores \citep{shi21}. Hot molecular cores are ideal targets for astrochemical studies because a variety of simple and complex organic molecules are frequently found towards these objects \citep{her09}. They are one of the earliest stages of star formation and play an important role in increasing the chemical complexity of the ISM \citep{shi21}. Hot molecular cores are small, compact objects ($\leq$0.1 pc) with a warm temperature ($\geq$100 K) and high gas density ($n_{\ce{H2}}$$\geq$10$^{6}$ cm$^{-3}$) that promote molecular evolution by thermal hopping on dust grains \citep{van98, wi14}. The lifetime of hot molecular cores is thought to be approximately 10$^{5}$ years (medium warm-up phase) to 10$^{6}$ years (slow warm-up phase) \citep{van98,vi04,gar06, gar08, gar13}.

 The hot molecular core candidate G358.93--0.03 MM1 is located in the high-mass star-formation region G358.93--0.03 at a distance of 6.75$\,{}\,^{\,+0.37}_{\,-0.68}$ kpc \citep{re14,bro19}. The total gas mass of G358.93--0.03 is 167$\pm$12\textup{M}$_{\odot}$ and its luminosity is $\sim$7.7$\times$10$^{3}$\textup{L}$_{\odot}$ \citep{bro19}. The high-mass star-formation region G358.93--0.03 contains eight sub-millimeter continuum sources, which are designated as G358.93--0.03 MM1 to G358.93--0.03 MM8 in order of decreasing right ascension \citep{bro19}. G358.93--0.03 MM1 is the brightest sub-millimeter continuum source that hosts a line-rich hot molecular core \citep{bro19, bay22}. Previously, maser lines of deuterated water (HDO), isocyanic acid (HNCO), and methanol (\ce{CH3OH}) were detected towards G358.93--0.03 MM1 using the ALMA, TMRT, and VLA radio telescopes \citep{bro19, chen20}. The rotational emission lines of methyl cyanide (\ce{CH3CN}) with transition J = 11(4)--10(4) were also detected from both G358.93--0.03 MM1 and G358.93--0.03 MM3 using the ALMA \citep{bro19}. The excitation temperature of \ce{CH3CN} towards the G358.93--0.03 MM1 and G358.93--0.03 MM3 is 172$\pm$3 K \citep{bro19}. The systematic velocities of G358.93--0.03 MM1 and G358.93--0.03 MM3 are --16.5$\pm$0.3 km s$^{-1}$ and --18.6$\pm$0.2 km s$^{-1}$, respectively \citep{bro19}. Recently,  rotational emission lines of the possible urea precursor molecule cyanamide (\ce{NH2CN}) were also detected towards G358.93--0.03 MM1 using the ALMA \citep{man23}.

In this article, we present the first detection of the simplest sugar-like molecule \ce{CH2OHCHO} towards the hot molecular core G358.93--0.03 MM1 using the Atacama Large Millimeter/Submillimeter Array (ALMA). ALMA data and their reductions are presented in Section~\ref{obs}. The line identification and the determination of the physical properties of the gas are presented in Section~\ref{res}. A discussion on the origin of \ce{CH2OHCHO} in this hot molecular core and conclusions are shown in Section~\ref{dis} and \ref{conclu}, respectively.

\begin{table*}
	%	\begin{minipage}[t]{\columnwidth}
	\centering
	%\scriptsize
	\caption{Summary of the continuum properties towards  G358.93--0.03 at wavelength 988 $\mu$m.}
	%	\begin{adjustbox}{width=0.5\textwidth}
	\begin{tabular}{ccccccccccccccccc}
		\hline 
		Source&R.A.&Decl.& Deconvolved source size &Integrated flux & Peak flux &RMS & Remark\\
		
		&     &     & ($^{\prime\prime}$$\times$$^{\prime\prime}$)&(mJy)&  (mJy beam$^{-1}$)&($\mu$Jy)&\\
		\hline
G358.93--0.03 MM1&17:43:10.1015 &--29:51:45.7057&0.116$\times$0.085&72.80$\pm$2.20&34.81$\pm$0.75&68.5&Resolved \\
\hline
G358.93--0.03 MM1A&17:43:10.0671&--29:51:46.4511&0.469$\times$0.411&67.71$\pm$2.60&3.53$\pm$0.13&21.1&Resolved\\

\hline

G358.93--0.03 MM2&17:43:10.0357&--29:51:44.9019&0.231$\times$0.088&14.20$\pm$1.40&~~4.47$\pm$0.35&35.5&Resolved\\
\hline

G358.93--0.03 MM2A&17:43:10.0209&--29:51:45.1577&--&2.30$\pm$0.58&1.52$\pm$0.18&20.3&Not resolved\\

\hline

G358.93--0.03 MM3&17:43:10.0145&--29:51:46.1933&0.072$\times$0.019&~~6.12$\pm$0.32&~~5.10$\pm$0.16&18.6&Resolved\\
		
\hline		
		
G358.93--0.03 MM4&17:43:09.9738&--29:51:46.0707&0.160$\times$0.087&10.50$\pm$2.42&~~4.38$\pm$0.32&49.8&Resolved\\
		
\hline		

G358.93--0.03 MM5&17:43:09.9063&--29.51.46.4814&0.245$\times$0.078&~~7.80$\pm$1.10&~~2.50$\pm$0.48&30.4&Resolved\\
\hline

G358.93--0.03 MM6&17:43:09.8962&--29:51:45.9802&0.107$\times$0.062&3.28$\pm$0.21&~~1.98$\pm$0.08&~7.9&Resolved\\
\hline

G358.93--0.03 MM7&17:43:09.8365&--29:51:45.9498&0.216$\times$0.116&13.91$\pm$1.82&~~3.85$\pm$0.39&38.5&Resolved\\
\hline		
		
G358.93--0.03 MM8&17:43:09.6761&--29:51:45.4688&--&~~6.50$\pm$0.11&~~4.582$\pm$0.06&~6.4&Not resolved\\
		\hline 
	\end{tabular}	
	%	\end{adjustbox}
	\label{tab:cont}
	%	\end{minipage}[t]{\columnwidth}
\end{table*}

\section{Observations and data reduction}
\label{obs}
The high-mass star-forming region G358.93--0.03 was observed using the ALMA band 7 receivers (PI: Crystal Brogan). The observation of G358.93--0.03 was performed on October 11, 2019, with a phase center of ($\alpha,\delta$)$_{\rm J2000}$ = (17:43:10.000, --29:51:46.000) and an on-source integration time of 756.0 sec. During the observations, a total of 47 antennas were used, with a minimum baseline of 14 m and a maximum baseline of 2517 m. J1550+0527 was used as the flux calibrator and bandpass calibrator, and J1744--3116 was used as the phase calibrator. The observed frequency ranges of G358.93--0.03 were 290.51--292.39 GHz, 292.49--294.37 GHz, 302.62--304.49 GHz, and 304.14--306.01 GHz, with a spectral resolution of 977 kHz (0.963 km s$^{-1}$).

We used the Common Astronomy Software Application ({\tt CASA 5.4.1}) for data reduction and imaging using the ALMA data reduction pipeline \citep{mc07}. We used the Perley-Butler 2017 flux calibrator model for flux calibration using the task {\tt SETJY} \citep{pal17}. After the initial data reduction using the CASA pipeline, we utilized task {\tt MSTRANSFORM} to separate the target data G358.93--0.03 with all the available rest frequencies. The continuum image is created by selecting line-free channels. Before creating the spectral images, the continuum emission is subtracted from the spectral data using the {\tt UVCONTSUB} task. To create the spectral images of G358.93--0.03, we used Briggs weighting \citep{bri95} and a robust value of 0.5. We used the CASA task {\tt IMPBCOR} to correct the synthesized beam pattern in continuum and spectral images.
	
\section{Results}
\label{res}
\subsection{Continuum emission}
\subsubsection{Sub-millimeter wavelength continuum emission towards  G358.93--0.03}
The continuum emission of G358.93--0.03 is observed at 303.39 GHz (988 $\mu$m), as shown in Figure~\ref{fig:cont}. The synthesized beam size is 0.412$^{\prime\prime}$$\times$0.363$^{\prime\prime}$. In the continuum emission image,  we observe eight sub-millimeter continuum sources, G358.93--0.03 MM1 to G358.93--0.03 MM8. Among the eight sources, G358.93--0.03 MM1 and G358.93--0.03 MM3 are known as hot molecular cores \citep{bro19}. Additionally, we also detected two other continuum sources associated with G358.93--0.03 MM1 and G358.93--0.03 MM2. We define these two continuum sources as G358.93--0.03 MM1A and G358.93--0.03 MM2A. We individually fit the 2D Gaussian for each source in G358.93--0.03 using the CASA task {\tt IMFIT} and estimate the integrated flux density, peak flux density, deconvolved source size, and RMS, which are shown in Table~\ref{tab:cont}. Except for G358.93--0.03 MM2A and G358.93--0.03 MM8, we see that the continuum emissions of other sources are resolved. \citet{bro19} also detected the sub-millimeter wavelength continuum emission from the eight individual continuum sources (G358.93--0.03 MM1 to G358.93--0.03 MM8) of G358.93--0.03 at frequencies of 195.58 GHz, 233.75 GHz, and 337.26 GHz.

 \begin{figure*}
	\centering
	\includegraphics[width=1.0\textwidth]{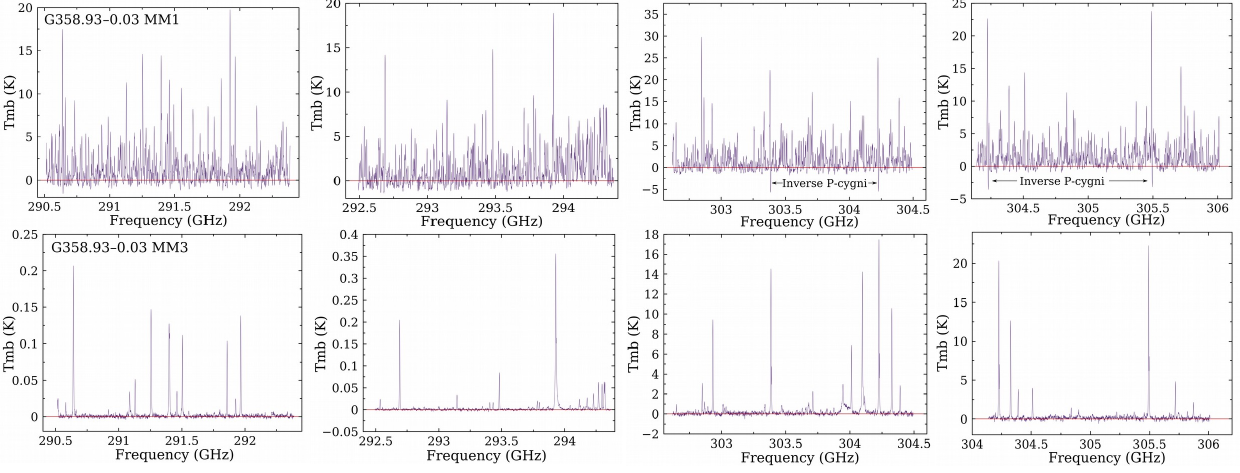}
	\caption{Sub-millimeter wavelength molecular emission spectra towards  G358.93--0.03 MM1 (upper panel) and G358.93--0.03 MM3 (lower panel), from ALMA band 7. The spectral resolution of the molecular spectra is 976.56 kHz. The red horizontal lines indicate the baseline of the molecular spectra.}
	\label{fig:fullspectra}
\end{figure*}

\begin{table}
	%	\begin{minipage}[t]{\columnwidth}
	\centering
	%\scriptsize
	\caption{Column densities of molecular \ce{H2} and optical depths towards the continuum sources in G358.93--0.03.}
		\begin{adjustbox}{width=0.48\textwidth}
	\begin{tabular}{ccccccccccccccccc}
		\hline 
Source&$N_{\ce{H2}}$&$T_{mb}$ &$\tau$\\
		
		     &(cm$^{-2}$)&(K) &\\
		\hline
		G358.93--0.03 MM1&(3.10$\pm$0.2)$\times$10$^{24}$&4.10 &27.72$\times$10$^{-3}$\\
		\hline
		G358.93--0.03 MM1A&(1.21$\pm$0.8)$\times$10$^{24}$ &0.41&13.71$\times$10$^{-3}$\\
		
		\hline
		
		G358.93--0.03 MM2&(1.53$\pm$0.7)$\times$10$^{24}$&0.52 &17.66$\times$10$^{-3}$\\
		\hline
		
		G358.93--0.03 MM2A&(5.22$\pm$0.2)$\times$10$^{23}$&0.17 &~~5.93$\times$10$^{-3}$\\
		
		\hline
		
		G358.93--0.03 MM3&(3.51$\pm$0.7)$\times$10$^{23}$&0.60 &~~4.01$\times$10$^{-3}$\\
		
		\hline		
		
		G358.93--0.03 MM4&(1.50$\pm$0.3)$\times$10$^{24}$&0.51&17.29$\times$10$^{-3}$\\
		
		\hline		
		
		G358.93--0.03 MM5&(8.59$\pm$0.5)$\times$10$^{23}$&0.29 &~~9.84$\times$10$^{-3}$\\
		\hline
		
		G358.93--0.03 MM6&(6.80$\pm$0.6)$\times$10$^{23}$&0.23&~~7.70$\times$10$^{-3}$\\
		\hline
		
		G358.93--0.03 MM7&(1.32$\pm$0.5)$\times$10$^{24}$ &0.45 &15.06$\times$10$^{-3}$\\
		\hline		
		
		G358.93--0.03 MM8&(1.57$\pm$0.8)$\times$10$^{24}$&0.54 &18.13$\times$10$^{-3}$\\
		\hline 
	\end{tabular}	
		\end{adjustbox}
	\label{tab:hydrogen}
	%	\end{minipage}[t]{\columnwidth}
\end{table}

\subsubsection{Estimation of molecular hydrogen (\ce{H2}) column density and optical depth ($\tau$) towards  G358.93--0.03}
\label{sec:hydrogen} 
Here we focus on estimating the molecular hydrogen column densities from all continuum sources in G358.93--0.03. The peak flux density ($S_\nu$) of the optically thin dust continuum emission can be expressed as
\begin{equation}
S_\nu = B_\nu(T_d)\tau_\nu\Omega_{beam}
\end{equation}
where the Planck function at dust temperature ($T_d$) is represented by $B_\nu(T_d)$ \citep{whi92}, $\tau_\nu$ is the optical depth, and $\Omega_{beam} = (\pi/4 \ln 2)\times \theta_{major} \times \theta_{minor}$ is the solid angle of the synthesized beam. The expression for the optical depth in terms of the mass density of dust can be written as,
\begin{equation}
\tau_\nu =\rho_d\kappa_\nu L
\end{equation}
where $\rho_d$ is the mass density of the dust, $\kappa_{\nu}$ is the mass absorption coefficient, and $L$  the path length. The mass density of the dust can be expressed in terms of the dust-to-gas mass ratio ($Z$),
\begin{equation}
\rho_d = Z\mu_H\rho_{H_2}=Z\mu_HN_{H_2}2m_H/L
\end{equation}
where $\mu_H$ is the mean atomic mass per hydrogen, $\rho_{H_2}$ is the hydrogen mass density, $m_H$ indicates the mass of hydrogen, and $N_{H_2}$ is the column density of hydrogen. For the dust temperature, $T_d$, we adopt 150 K, as derived by \citet{chen20}  for the two hot cores, G358.93--0.03 MM1 and G358.93--0.03 MM3. For the rest of the cores, we adopt a dust temperature of 30.1 K as estimated by \citet{ste21}. We also take $\mu_H = 1.41$ and $Z = 0.01$ \citep{cox00}. The peak flux density of all the continuum sources in G358.93--0.03 at frequency 303.39 GHz is listed in Table~\ref{tab:cont}. From equations 1, 2, and 3, the column density of molecular hydrogen can be expressed as,

\begin{equation}
N_{H_2} = \frac{S_\nu /\Omega}{2\kappa_\nu B_\nu(T_d)Z\mu_H m_H}
\end{equation}
For the estimation of the mass absorption coefficient ($\kappa_{\nu}$), we use the formula $\kappa_\nu = 0.90(\nu /230 ~\textrm{GHz})^{\beta}$~ cm$^{2}$~g$^{-1}$ \citep{moto19}, where $k_{230} = 0.90$~cm$^{2}$~g$^{-1}$ indicates the emissivity of the dust grains at a gas density of $\rm{10^{6}\ cm^{-3}}$. We use the dust spectral index $\beta$ $\sim$ 1.7 \citep{bro19}. From equation 4, we find the column densities of molecular hydrogen ($N_{\ce{H2}}$) towards all observed continuum sources in G358.93--0.03, which we report in Table~\ref{tab:hydrogen}.

We also determine the value of the dust optical depth ($\tau_\nu$) using the following equation,

\begin{equation}
T_{mb} = T_{d}(1-exp(-\tau_\nu))
\end{equation}
where $T_{mb}$ represents the brightness temperature and $T_{d}$ is the dust temperature. For an estimation of the brightness temperature ($T_{mb}$), we use the Rayleigh-Jeans approximation, 1 Jy beam$^{-1} \equiv$ 118 K. Using equation 5, we estimate the dust optical depth towards all observed continuum sources in G358.93--0.03 (listed in Table~\ref{tab:hydrogen}). We find that the optical depth of all observed continuum sources is less than 1, indicating that all observed continuum sources in G358.93--0.03 are optically thin at the frequency of 303.39 GHz.

\begin{figure*}
	\centering
	\includegraphics[width=0.91\textwidth]{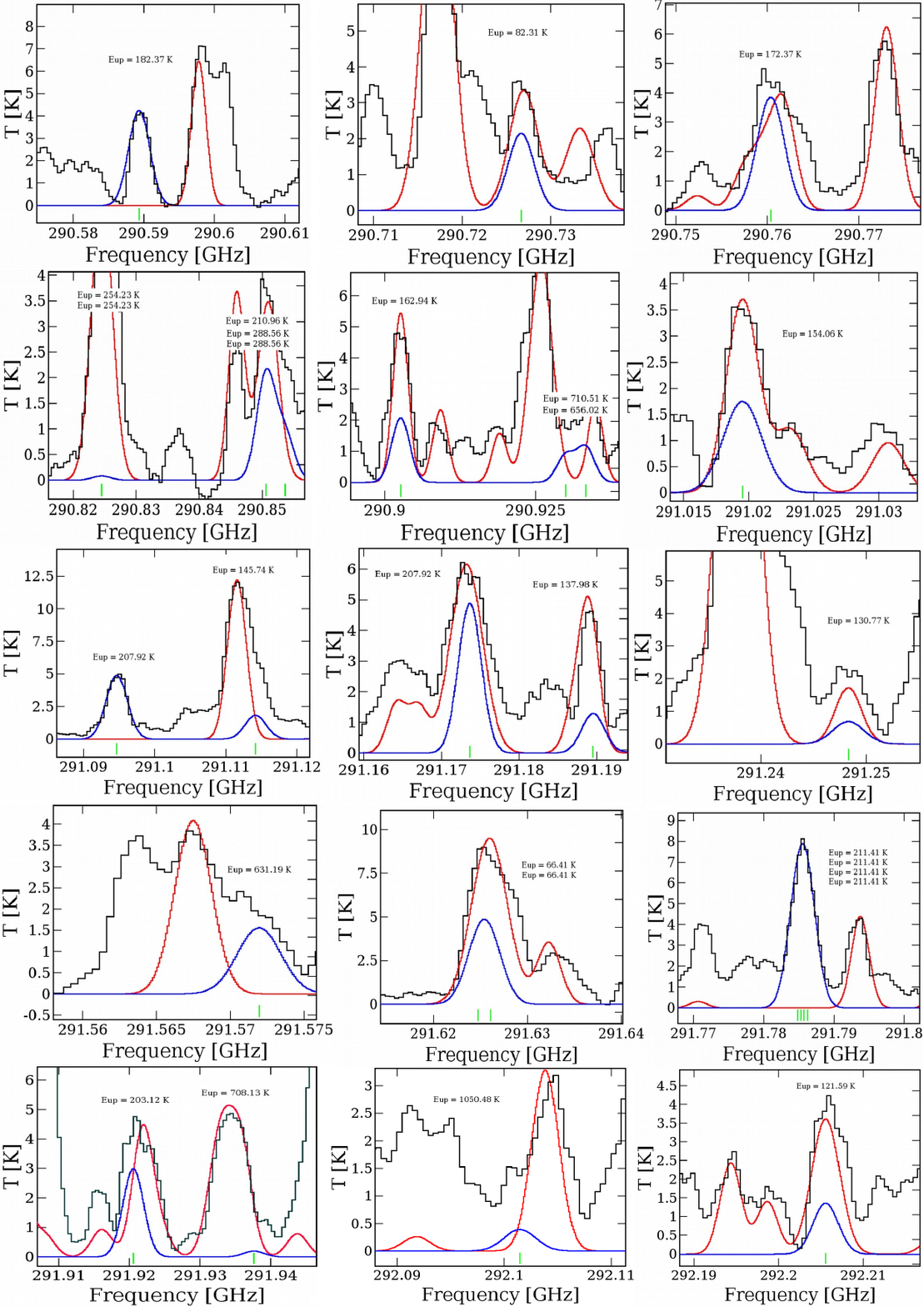}
	\caption{Rotational emission lines of \ce{CH2OHCHO} towards the hot molecular core G358.93--0.03 MM1. The black spectra are the observations spectra of G358.93--0.03 MM1. The blue spectra represent the LTE model spectrum of just \ce{CH2OHCHO}, while the red spectra are the LTE model spectra, including all species. The green vertical lines in the LTE spectra indicate the rest frequency positions of the detected transitions of  \ce{CH2OHCHO}.}
	\label{fig:ltespec}
\end{figure*}

\begin{figure*}
	\text{{\large Figure~3 Continued.}}
	\centering
	\includegraphics[width=0.93\textwidth]{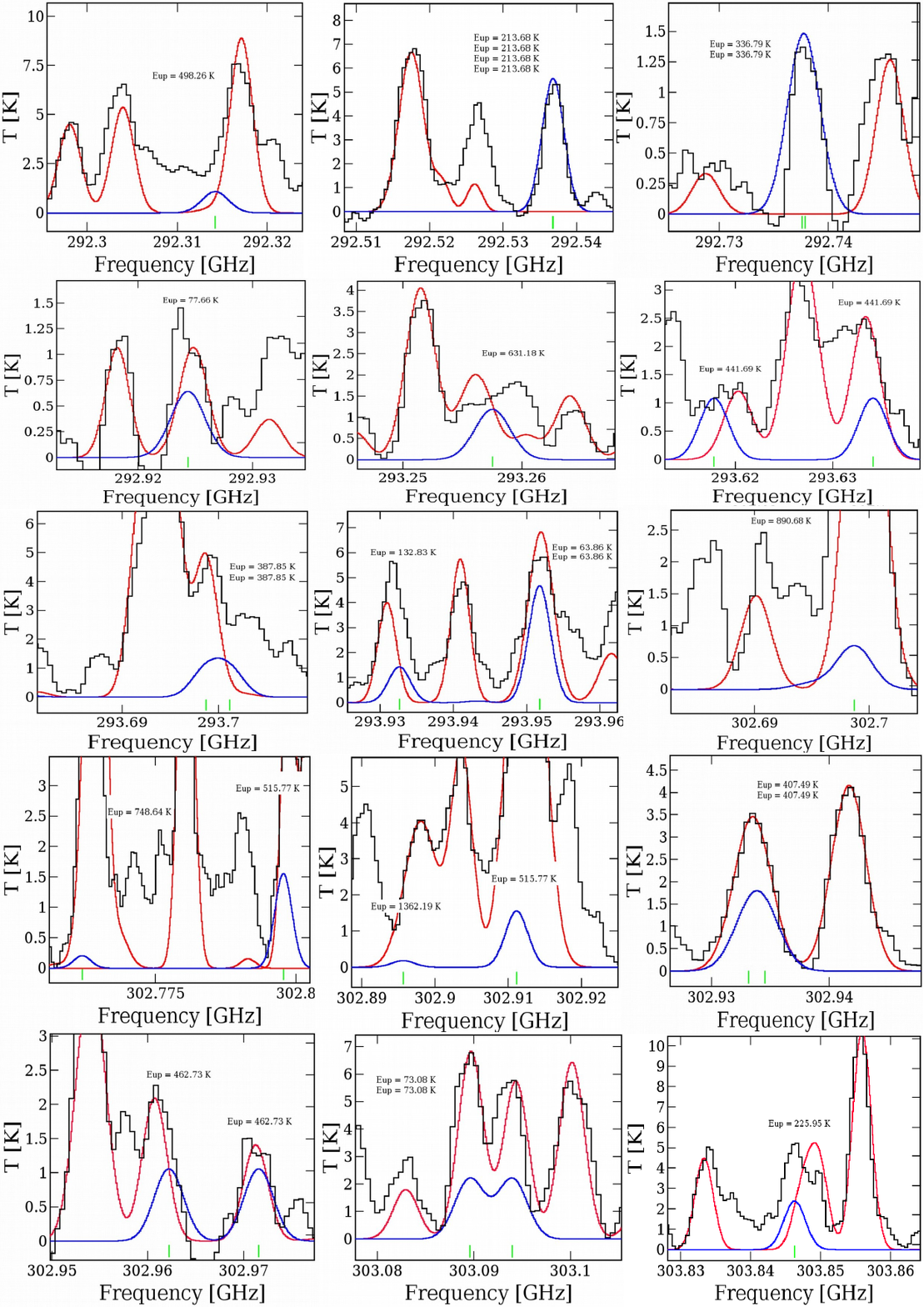}
\end{figure*}

\begin{figure*}
	\text{{\large Figure~3 Continued.}}
	\centering
	\includegraphics[width=0.93\textwidth]{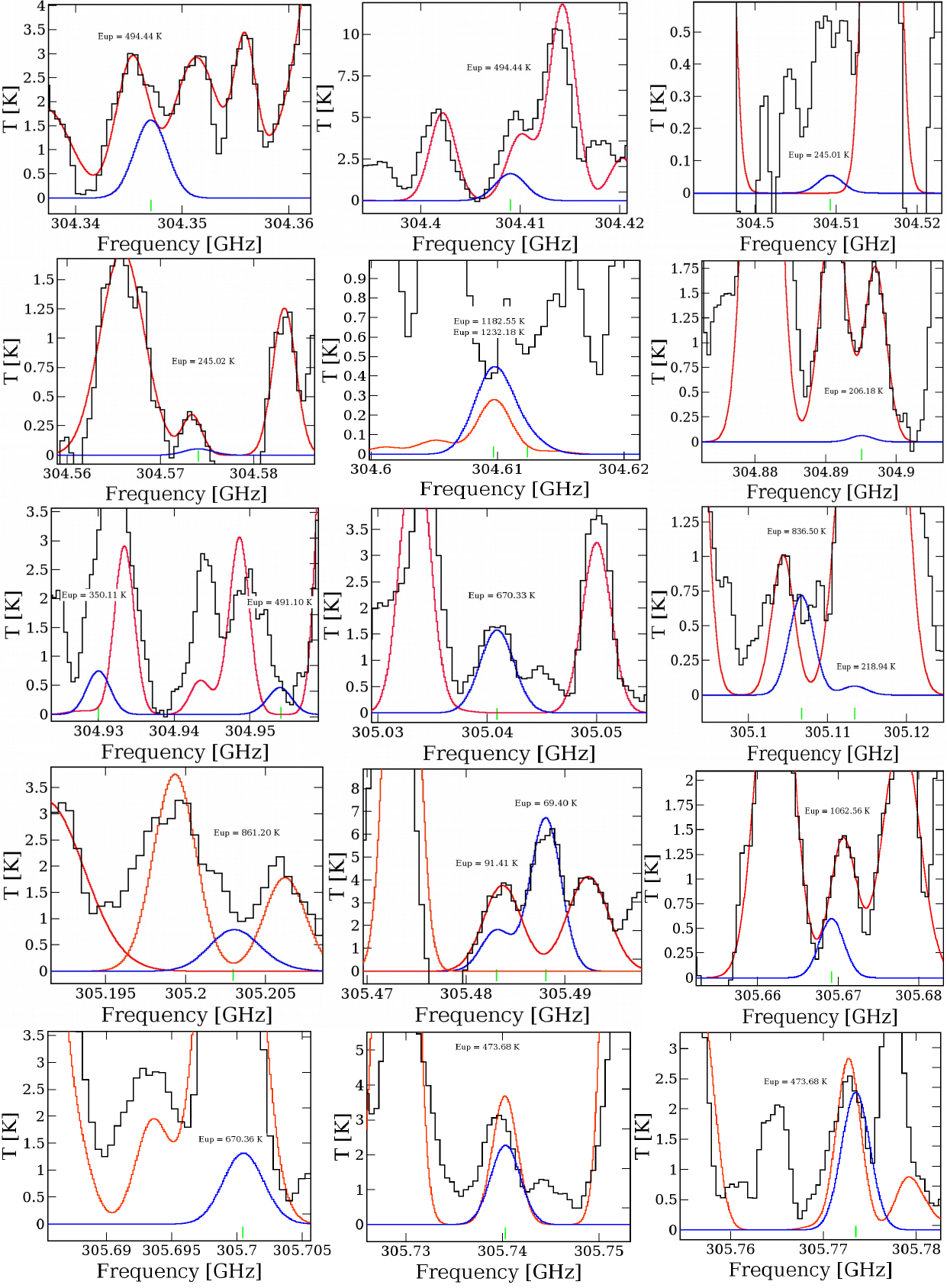}
\end{figure*}

\begin{table*}
	%\begin{minipage}[t]{\columnwidth}
	\centering
	\scriptsize 
	\caption{Summary of the line parameters of  \ce{CH2OHCHO} towards  G358.93--0.03 MM1.}
	\begin{adjustbox}{width=1.0\textwidth}
		\begin{tabular}{ccccccccccccccccc}
			\hline 
			Observed frequency &Transition & $E_{u}$ & $A_{ij}$ &$G_{up}$&FWHM &Optical depth&$I_p$$^{\dagger}$ &$\rm{\int T_{mb}dV}$ &Remark\\
			
			(GHz) &(${\rm J^{'}_{K_a^{'}K_c^{'}}}$--${\rm J^{''}_{K_a^{''}K_c^{''}}}$) &(K)&(s$^{-1}$) &&(km s$^{-1}$) &($\tau$)&(K) &(K.km s$^{-1}$) & \\
			\hline
			~~290.589$^{*}$&18(12,6)--18(11,7)&182.3&2.59$\times$10$^{-4}$&37&3.2$\pm$0.2 &1.2$\times$10$^{-2}$&4.1&11.9$\pm$1.2 &Non blended\\
			
			290.726&15(5,10)--14(4,11)&~~82.3& 2.23$\times$10$^{-4}$&31&--&1.3$\times$10$^{-2}$&3.6&--&Blended with DCOOH\\
			
			~~290.760$^{*}$&17(12,6)--17(11,7)&172.3&2.39$\times$10$^{-4}$&35&--&1.0$\times$10$^{-2}$&4.8&--&Blended with \ce{CH3O}$^{13}$CHO\\
			
			290.824&25(11,15)--24(11,14)&254.2&7.78$\times$10$^{-6}$&51&--&3.9$\times$10$^{-4}$&5.7&--&Blended with \ce{CH3OCHO}\\
			
			290.824&25(11,14)--24(11,13)&254.2&7.78$\times$10$^{-6}$&51&--&3.9$\times$10$^{-4}$ &5.7&--&Blended with \ce{CH3OCHO}\\
			
			290.850&26(4,22)--25(5,21)&210.9&3.16$\times$10$^{-4}$&53&--&1.9$\times$10$^{-2}$ &3.7&--&Blended with \ce{CH3COOH}\\
			
			290.853&33(1,32)--33(0,33)&288.5&6.82$\times$10$^{-5}$&67&--&4.0$\times$10$^{-3}$ &2.5&--&Blended with \ce{H2CCNH}\\
			
			290.853&33(2,32)--33(1,33)&288.5&6.82$\times$10$^{-5}$&67&--&4.0$\times$10$^{-3}$&2.5&--&Blended with \ce{H2CCNH}\\
			
			~~290.902$^{*}$&16(12,4)--16(11,5)&162.9&2.16$\times$10$^{-4}$&33&--&9.7$\times$10$^{-3}$&4.9&--&Blended with \ce{CH3COOH}\\
			
			290.929&48(9,40)--48(8,41)&710.5&3.81$\times$10$^{-4}$&97&--&8.0$\times$10$^{-3}$ &1.8&--&Blended with \ce{CH3CH2CN}\\
			
			290.933&44(13,32)--44(12,33)&656.0&4.49$\times$10$^{-4}$&89&--&8.2$\times$10$^{-3}$ &2.8&--&Blended with \ce{CH3CH2CN}\\
			
			~~291.019$^{*}$&15(12,4)--15(11,5)&154.0&1.89$\times$10$^{-4}$&31&--&8.1$\times$10$^{-3}$ &3.5&--&Blended with \ce{CH3COOH}\\
			
			291.094&27(2,25)--26(3,24)&207.9&5.47$\times$10$^{-4}$&55&3.3$\pm$0.3&3.4$\times$10$^{-2}$ &5.0&14.3$\pm$2.3&Non blended\\
			
			~~291.114$^{*}$&14(12,2)--14(11,3)&145.7&1.55$\times$10$^{-4}$&29&--&6.5$\times$10$^{-3}$ &12.0&--&Blended with \ce{CH3}$^{18}$OH\\
			
			291.173&27(3,25)--26(2,24)&207.9&5.47$\times$10$^{-4}$&55&--&3.5$\times$10$^{-2}$&5.7&--&Blended with \ce{CH3CCH}\\
			
			~~291.189$^{*}$&13(12,2)--13(11,3)&137.9&1.15$\times$10$^{-4}$&27&--&4.5$\times$10$^{-3}$&4.5&--&Blended with \ce{CH3COCH3}\\
			
			~~291.248$^{*}$&12(12,0)--12(11,1)&130.7&6.40$\times$10$^{-5}$&25&--&2.5$\times$10$^{-3}$&2.3&--&Blended with HC$^{18}$O\ce{NH2}\\
			
			291.571&43(13,30)--43(12,31)&631.1&4.47$\times$10$^{-4}$&87&--&1.1$\times$10$^{-2}$&2.4&--&Blended with $^{13}$\ce{CH3CH2CN}\\
			
			291.624&11(7,5)--10(6,4)&~~66.4&4.49$\times$10$^{-4}$&23&--&2.0$\times$10$^{-2}$ &8.9&--&Blended with \ce{DCONH2}\\
			
			291.626&11(7,4)--10(6,5)&~~66.4&4.49$\times$10$^{-4}$&23&--&2.0$\times$10$^{-2}$ &8.9&--&Blended with \ce{DCONH2}\\
			
			291.784&28(1,27)--27(2,26)&211.4&6.46$\times$10$^{-4}$&57&3.4$\pm$0.1&4.2$\times$10$^{-2}$ &8.1&27.9$\pm$4.8&Non blended\\
			
			291.784&28(2,27)--27(2,26)&211.4&9.58$\times$10$^{-6}$&57&3.4$\pm$0.1&4.2$\times$10$^{-2}$&8.1&27.9$\pm$4.8&Non blended\\
			
			291.784&28(1,27)--27(1,26)&211.4&9.58$\times$10$^{-6}$&57&3.4$\pm$0.1&4.2$\times$10$^{-2}$&8.1&27.9$\pm$4.8&Non blended\\
			
			291.786&28(2,27)--27(1,26)&211.4&6.46$\times$10$^{-4}$&57&3.4$\pm$0.1&4.2$\times$10$^{-2}$&8.1&27.9$\pm$4.9&Non blended\\
			
			291.920&26(4,23)--25(3,22)&203.1&4.41$\times$10$^{-4}$&53&--&2.8$\times$10$^{-2}$ &4.7&--&Blended with $^{13}$\ce{CH2CHCN}\\
			
			291.937&47(11,37)--46(12,34)&708.1&8.68$\times$10$^{-5}$&95&--&1.8$\times$10$^{-3}$ &4.8&--&Blended with \ce{CH3COOH}\\
			
			292.101&58(11,47)--58(10,48)&1050.4&4.50$\times$10$^{-4}$&117&--&3.7$\times$10$^{-3}$ &1.5&--&Blended with \ce{C2H5OH}\\
			
			292.205&19(5,15)--18(4,14)&121.5&2.13$\times$10$^{-4}$&39&--&1.2$\times$10$^{-2}$ &4.2&--&Blended with E-\ce{CH3CHO}\\
			
			292.314&41(5,36)--41(4,37)&498.2&2.79$\times$10$^{-4}$&83&--&1.1$\times$10$^{-2}$ &7.5&--&Blended with \ce{CH3OCHO}\\
			
			292.536&29(0,29)--28(1,28)&213.6&7.37$\times$10$^{-4}$&59&3.3$\pm$0.2&5.0$\times$10$^{-3}$&5.3&15.4$\pm$3.5&Non blended\\
			
			292.536&29(1,29)--28(1,28)&213.6&9.77$\times$10$^{-6}$&59&3.3$\pm$0.2&5.0$\times$10$^{-3}$&5.3&15.4$\pm$3.5&Non blended\\
			
			292.536&29(0,29)--28(0,28)&213.6&9.77$\times$10$^{-6}$&59&3.3$\pm$0.2&5.0$\times$10$^{-3}$&5.3&15.4$\pm$3.5&Non blended\\
			
			292.536&29(1,29)--28(0,28)&213.6&7.37$\times$10$^{-4}$&59&3.3$\pm$0.2&5.0$\times$10$^{-3}$&5.3&15.4$\pm$3.5&Non blended\\
			
			292.737&35(2,33)--35(1,34)&336.7&1.31$\times$10$^{-4}$&71&3.2$\pm$0.5&7.0$\times$10$^{-3}$&1.3&3.5$\pm$0.1&Non blended\\
			
			292.737&35(3,33)--35(2,34)&336.7&1.31$\times$10$^{-4}$&71&3.2$\pm$0.6&7.0$\times$10$^{-3}$&1.3&3.5$\pm$0.1&Non blended\\
			
			292.924&15(4,11)--14(3,12)&~~77.6&1.09$\times$10$^{-4}$&31&--&6.1$\times$10$^{-3}$ &1.4&--&Blended with \ce{CH3C3N}\\
			
			293.257&43(13,31)--43(12,32)&631.1&4.54$\times$10$^{-4}$&87&--&1.1$\times$10$^{-2}$&1.5&--&Blended with \ce{HC3N}\\
			
			293.617&39(4,35)--39(3,36)&441.7&2.37$\times$10$^{-4}$&79&--&9.8$\times$10$^{-3}$ &1.0&--&Blended with $^{13}$\ce{CH3CH2CN}\\
			
			293.634&39(5,35)--39(4,36)&441.7&2.37$\times$10$^{-4}$&79&--&9.8$\times$10$^{-3}$ &2.5&--&Blended with \ce{CH3OCH3}\\
			
			293.698&37(3,34)--37(2,35)&387.8&1.88$\times$10$^{-4}$&75&--&8.9$\times$10$^{-3}$ &5.0&--&Blended with \ce{CH3COOH}\\
			
			293.701&37(4,34)--37(3,35)&387.8&1.88$\times$10$^{-4}$&75&--&8.9$\times$10$^{-3}$ &5.0&--&Blended with \ce{CH3COOH}\\
			
			293.932&20(5,16)--19(4,15)&132.8&2.21$\times$10$^{-4}$&41&--&1.3$\times$10$^{-2}$ &5.6&--&Blended with \ce{CH2CH}$^{13}$CN\\
			
			293.951&9(8,2)--8(7,3)&~~63.8&6.30$\times$10$^{-4}$&19&--&2.2$\times$10$^{-2}$ &5.7&--&Blended with $^{13}$\ce{CH3CH2CN}\\
			
			293.951&9(8,1)--8(7,2)&~~63.8&6.30$\times$10$^{-4}$&19&--&2.2$\times$10$^{-2}$&5.7&--&Blended with $^{13}$\ce{CH3CH2CN}\\
			
			302.698&52(14,39)--52(13,40)&890.6&5.30$\times$10$^{-4}$&105&--&6.1$\times$10$^{-3}$&6.5&--&Blended with \ce{CH2DOH}\\
			
			302.761&48(12,36)--47(13,35)&748.6&1.07$\times$10$^{-4}$&97&--&1.8$\times$10$^{-3}$&6.3&--&Blended with HC$^{18}$O\ce{NH2}\\
			
			302.797&38(13,25)--38(12,36)&515.7&4.70$\times$10$^{-4}$&77&--&1.3$\times$10$^{-2}$&5.3&--&Blended with \ce{CH3SH}\\
			
			302.895&66(13,53)--66(12,54)&1362.2&5.32$\times$10$^{-4}$&133&--&1.6$\times$10$^{-3}$&4.0&--&Blended with \ce{CH3O}$^{13}$CHO\\
			
			302.911&38(13,26)--38(12,27)&515.7&4.71$\times$10$^{-4}$&77&--&1.2$\times$10$^{-2}$&15.3&--&Blended with \ce{CH3OH}\\
			
			302.933&38(3,35)--38(2,36)&407.5&2.09$\times$10$^{-4}$&77&--&8.5$\times$10$^{-3}$&3.5&--&Blended with H$^{13}$CCCN\\
			
			302.934&38(4,35)--38(3,36)&407.5&2.09$\times$10$^{-4}$&77&--&8.5$\times$10$^{-3}$&3.5&--&Blended with H$^{13}$CCCN\\
			
			302.962&40(4,36)--40(3,37)&462.7&2.54$\times$10$^{-4}$&81&--&9.5$\times$10$^{-3}$&2.2&--&Blended with \ce{CH2}CH$^{13}$CN\\
			
			302.971&40(5,36)--40(4,37)&462.7&2.54$\times$10$^{-4}$&81&--&9.5$\times$10$^{-3}$&1.5&--&Blended with HCOOD\\
			
			303.089&12(7,6)--11(6,5)&~~73.0&4.68$\times$10$^{-4}$&25&--&2.1$\times$10$^{-2}$&7.2&--&Blended with \ce{NH2CO}$^{+}$\\
			
			303.094&12(7,5)--11(6,6)&~~73.0&4.68$\times$10$^{-4}$&25&--&2.0$\times$10$^{-2}$&6.8&--&Blended with c-HCOOH\\
			
			303.846&27(4,23)--26(5,22)&225.9&3.86$\times$10$^{-4}$&55&--&2.1$\times$10$^{-2}$&5.2&--&Blended with D$^{13}$CCCN\\
			
			304.347&37(13,24)--37(12,25)&494.4&4.71$\times$10$^{-4}$&75&--&1.4$\times$10$^{-2}$&3.3&--&Blended with \ce{CH3CDO}\\
			
			304.409&37(13,25)--37(12,26)&494.4&4.71$\times$10$^{-4}$&75&--&1.4$\times$10$^{-2}$&5.3&--&Blended with NCHCCO\\
			
			304.509&26(9,18)--25(9,16)&245.0&9.75$\times$10$^{-6}$&53&--&4.8$\times$10$^{-4}$&0.5&--&Blended with $^{13}$\ce{CH3CN}\\
			
			304.574&26(9,17)--25(9,16)&245.0&9.76$\times$10$^{-6}$&53&--&4.8$\times$10$^{-4}$&0.3&--&Blended with \ce{C2H5}C$^{15}$N\\
			
			304.609&61(14,48)--61(13,49)&1182.5&5.58$\times$10$^{-4}$&123&--&2.8$\times$10$^{-3}$&0.8&--&Blended with \ce{C2H5CN}\\
			
			304.612&61(17,45)--60(18,42)&1232.1&1.07$\times$10$^{-4}$&123&--&4.6$\times$10$^{-4}$&0.8&--&Blended with \ce{C2H5CN}\\

			304.895&25(6,19)--24(6,18)&206.1&1.05$\times$10$^{-5}$&51&--&5.7$\times$10$^{-4}$&0.9&--&Blended with H$^{13}$CCN\\
			
			304.930&33(7,26)--32(8,25)&350.1&1.50$\times$10$^{-4}$&67&--&6.6$\times$10$^{-3}$&4.9&--&Blended with \ce{CH3OCHO}\\			
			
			304.954&39(9,30)--38(10,29)&491.1&1.17$\times$10$^{-4}$&79&--&3.8$\times$10$^{-3}$&1.8&--&Blended with \ce{HC(O)NH2}\\
			
			305.040&47(7,40)--47(6,41)&670.3&3.90$\times$10$^{-4}$&95&3.4$\pm$0.6  &6.1$\times$10$^{-3}$&1.6&5.806$\pm$1.87&Non blended\\

			305.106&52(10,43)--52(9,44)&836.5&4.59$\times$10$^{-4}$&105&--&3.1$\times$10$^{-3}$&0.7&--&Blended with \ce{CH3COOH}\\
			
			305.113&26(6,21)--25(6,20)&218.9&1.05$\times$10$^{-5}$&53&--&2.6$\times$10$^{-4}$&3.0&--&Blended with \ce{C2H5CN}\\

			\hline
		\end{tabular}	
	\end{adjustbox}
	\label{tab:MOLECULAR DATA}\\
	%\label{tab:MOLECULAR DATA}\\
	%{\color{blue}{*}}--The transition of t-HC(O)SH contain double with frequency difference  $\leq$100 kHz. The second transition is not shown.\\
	%		\end{minipage}[t]{\columnwidth}
\end{table*}

\begin{table*}
	%\begin{minipage}[t]{\columnwidth}
	\centering
	\scriptsize 
	\text{{\large Table~2 Continued.}}
	%\caption{Summary of the LTE fitted line parameters of the t-HC(O)SH towards the IRAS 16293 B.}
	\begin{adjustbox}{width=1.0\textwidth}
		\begin{tabular}{ccccccccccccccccc}
			\hline 
			Observed frequency &Transition & $E_{u}$ & $A_{ij}$ &$G_{up}$&FWHM &Optical depth&$I_p$$^{\dagger}$ &$\rm{\int T_{mb}dV}$ &Remark\\
			
			(GHz) &(${\rm J^{'}_{K_a^{'}K_c^{'}}}$--${\rm J^{''}_{K_a^{''}K_c^{''}}}$) &(K)&(s$^{-1}$) &&(km s$^{-1}$) &($\tau$)&(K) &(K~km s$^{-1}$) & \\
			\hline
			305.203&51(14,38)--51(13,39)&861.2&5.38$\times$10$^{-4}$&103&--&3.1$\times$10$^{-3}$&3.2&--&Blended with \ce{CH2DCHO}\\

			305.483&16(5,11)--15(4,12)&~~91.4&2.29$\times$10$^{-4}$&33&--&5.4$\times$10$^{-3}$&3.9&--&Blended with \ce{CH3COOH}\\
			
			305.488$^{*}$&10(8,3)--9(7,2)&~~69.4&6.41$\times$10$^{-4}$&21&3.4$\pm$0.5 &1.1$\times$10$^{-2}$&6.2&24.2$\pm$2.6 &Non blended\\
			
			305.669&57(15,42)--57(14,43)&1062.5&5.56$\times$10$^{-4}$&115&--&1.9$\times$10$^{-3}$&1.4&--&Blended with \ce{CH3COOH}\\
			
			305.700&47(8,40)--47(7,41)&670.3&3.93$\times$10$^{-4}$&95&--&4.1$\times$10$^{-3}$&15.6&--&Blended with \ce{CH2DCHO}\\
			
			305.740&36(13,23)--36(12,24)&473.6&4.71$\times$10$^{-4}$&73&--&7.0$\times$10$^{-3}$&3.5&--&Blended with \ce{CH3COOH}\\
			
			305.773&36(13,24)--36(12,25)&473.6&4.71$\times$10$^{-4}$&73&--&7.0$\times$10$^{-3}$&2.5&--&Blended with \ce{CH3}$^{18}$OH\\	
			
			\hline
		\end{tabular}	
	\end{adjustbox}
	{{*}}-- There are two transitions that have close frequencies ($\leq  100$ kHz), and only the frequency of the first transition is shown.\\
	$\dagger$--$I_p$ is the peak intensity of the emission lines of \ce{CH2OHCHO}. 
	%		\end{minipage}[t]{\columnwidth}
\end{table*}

\subsection{Line emission from G358.93--0.03}
From the spectral images of G358.93--0.03, we see that only the spectra of G358.93--0.03 MM1 and G358.93--0.03 MM3 show any line emission. The synthesized beam sizes of the spectral images of G358.93--0.03 at frequency ranges of 290.51--292.39 GHz, 292.49--294.37 GHz, 302.62--304.49 GHz, and 304.14--306.01 GHz are 0.425$^{\prime\prime}\times$0.369$^{\prime\prime}$, 0.427$^{\prime\prime}\times$0.376$^{\prime\prime}$, 0.413$^{\prime\prime}\times$0.364$^{\prime\prime}$, and 0.410$^{\prime\prime}\times$0.358$^{\prime\prime}$, respectively. We extract the molecular spectra from G358.93--0.03 MM1 and G358.93--0.03 MM3 by drawing a 0.912$^{\prime\prime}$ diameter circular region, which is larger than the line emitting regions of G358.93--0.03 MM1 and G358.93--0.03 MM3. The phase centre of G358.93--0.03 MM1 is RA (J2000) = 17$^{h}$43$^{m}$10$^{s}$.101, Dec (J2000) = --29$^\circ$51$^{\prime}$45$^{\prime\prime}$.693. The phase centre of G358.93--0.03 MM3 is RA (J2000) = 17$^{h}$43$^{m}$10$^{s}$.0144, Dec (J2000) = --29$^\circ$51$^{\prime}$46$^{\prime\prime}$.193. The resultant spectra of G358.93--0.03 MM1 and G358.93--0.03 MM3 are shown in Figure~\ref{fig:fullspectra}. From the spectra, it can be seen that G358.93--0.03 MM1 is more chemically rich than G358.93--0.03 MM3. Additionally, we also observe the signature of an inverse P-Cygni profile associated with the \ce{CH3OH} emission lines towards  G358.93--0.03 MM1. This may indicate that the hot molecular core G358.93--0.03 MM1 is undergoing infall. We do not observe any evidence of an inverse P-Cygni profile in the spectra of G358.93--0.03 MM3. The systematic velocities ($V_{LSR}$) of G358.93--0.03 MM1 and G358.93--0.03 MM3 are --16.5 km s$^{-1}$ and --18.2 km s$^{-1}$, respectively, \citep{bro19}.

\subsubsection{Identification of \ce{CH2OHCHO} towards  G358.93--0.03 MM1}
\label{sec:fitting}
To identify the rotational emission lines of \ce{CH2OHCHO}, we assume local thermodynamic equilibrium (LTE) and use the Cologne Database for Molecular Spectroscopy (CDMS) \citep{mu05}. For LTE modelling, we use CASSIS \citep{vas15}. The LTE assumption is valid in the inner region of G358.93--0.03 MM1 because the gas density of the warm inner region of the hot core is 2$\times$10$^{7}$ cm$^{-3}$ \citep{ste21}. To fit the LTE model spectra of \ce{CH2OHCHO} over the observed molecular spectra, we use the Markov Chain Monte Carlo (MCMC) algorithm in CASSIS. Previously, \citet{gor20} discussed the fitting of the LTE model spectrum using MCMC in detail. We have identified a total of seventy-five transitions of \ce{CH2OHCHO} towards G358.93--0.03 MM1 between the frequency ranges of 290.51--292.39 GHz, 292.49--294.37 GHz, 302.62--304.49 GHz, and 304.14--306.01 GHz. The upper-level energies of the identified seventy-five transitions of \ce{CH2OHCHO} vary from 63.86 K to 1362.19 K. Among the detected seventy-five transitions, we find that only fourteen transitions of \ce{CH2OHCHO} are not blended, and these lines are identified to be higher than 4$\sigma$ (confirmed from LTE modelling). The upper-level energies of the non-blended transitions of \ce{CH2OHCHO} vary between 69.40 K and 670.33 K. There are no missing transitions in \ce{CH2OHCHO} in the observed frequency ranges. The blended transitions of \ce{CH2OHCHO} will not be considered in our modelling. From the LTE modelling, the best-fit column density of \ce{CH2OHCHO} is found to be (1.52$\pm$0.9)$\times$10$^{16}$ cm$^{-2}$ with an excitation temperature of 300$\pm$68.5 K and a source size of 0.45$^{\prime\prime}$. The FWHM of the LTE spectra of \ce{CH2OHCHO} is 3.35 km s$^{-1}$. We observed that the FWHM of the spectra of \ce{CH2OHCHO} is nearly similar to the FWHM of another molecule \ce{CH3CN} towards G358.93--0.03 MM1, which was estimated by \citet{bro19}. The LTE-fitted rotational emission spectra of \ce{CH2OHCHO} are shown in Figure~\ref{fig:ltespec}. In addition to \ce{CH2OHCHO}, the hot molecular core G358.93--0.03 MM1 also contains several other complex organic molecules, including \ce{CH3OCHO}, \ce{CH3COOH}, \ce{CH3NH2}, \ce{CH3OH}, \ce{CH3SH}, \ce{C2H5CN}, and \ce{C2H3CN}, which we discuss in a separate paper.
 
We report the details of all the detected \ce{CH2OHCHO} lines in Table~\ref{tab:MOLECULAR DATA}. Additionally, we also fitted the Gaussian model over the non-blended emission lines of \ce{CH2OHCHO} to estimate the proper full width-half maximum (FWHM) in km s$^{-1}$ and integrated intensity ($\rm{\int T_{mb}dV}$) in K~km s$^{-1}$.  We have observed three different non-blended emission lines of \ce{CH2OHCHO} at frequencies of 291.784 GHz, 292.536 GHz, and 292.737 GHz that contain multiple transitions of \ce{CH2OHCHO}. These transitions are reported in Table~\ref{tab:MOLECULAR DATA}. We cannot separate these observed transitions as they are very close to each other, i.e., blended with each other. To obtain the line parameters of those transitions of \ce{CH2OHCHO}, we have fitted a multiple-component Gaussian using the Levenberg-Marquardt algorithm in CASSIS to the observed spectra. For multiple Gaussian fittings, we have used fixed values of velocity separation and the expected line intensity ratio. During the fitting of a multi-component Gaussian, only the FWHM is kept as a free parameter. This method works well in the observed spectral profiles around 291.784 GHz, 292.536 GHz, and 292.737 GHz of \ce{CH2OHCHO}. The summary of the detected transitions and spectral line properties of \ce{CH2OHCHO} is presented in Table~\ref{tab:MOLECULAR DATA}.

To determine the fractional abundance of \ce{CH2OHCHO}, we use the column density of \ce{CH2OHCHO} inside the 0.45$^{\prime\prime}$ beam, and divide it by the \ce{H2} column density found in Section~\ref{sec:hydrogen}. The fractional abundance of \ce{CH2OHCHO} with respect to \ce{H2} towards the G358.93--0.03 MM1 is (4.90$\pm$2.92)$\times$10$^{-9}$, where the column density of \ce{H2} towards the G358.93--0.03 MM1 is (3.10$\pm$0.2)$\times$10$^{24}$ cm$^{-2}$. Recently, \citet{min20} found that the abundance of \ce{CH2OHCHO} towards another hot molecular core, G31.41+0.31, was (5.0$\pm$1.4)$\times$10$^{-9}$, which is close to our derived abundance of \ce{CH2OHCHO} towards G358.93--0.03 MM1. This indicates that the chemical formation route(s) of \ce{CH2OHCHO} towards the G358.93--0.03 MM1 may be similar to those in G31.41+0.31.

\begin{figure*}
	\centering
	\includegraphics[width=1.0\textwidth]{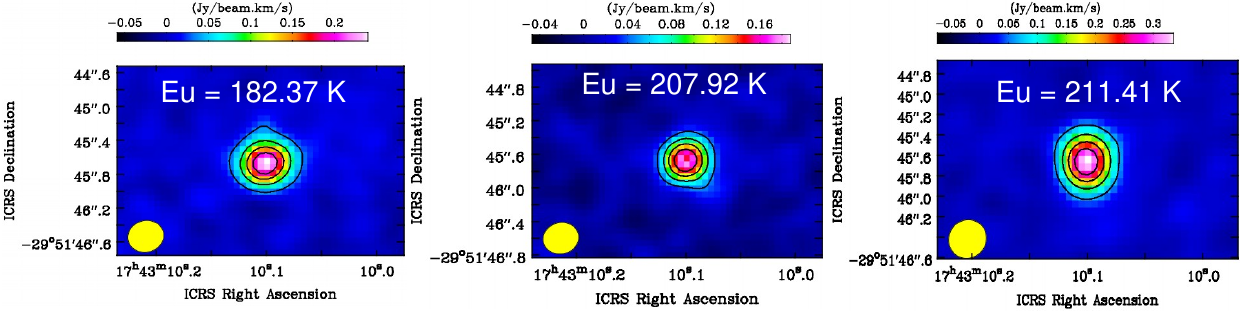}
	\includegraphics[width=1.0\textwidth]{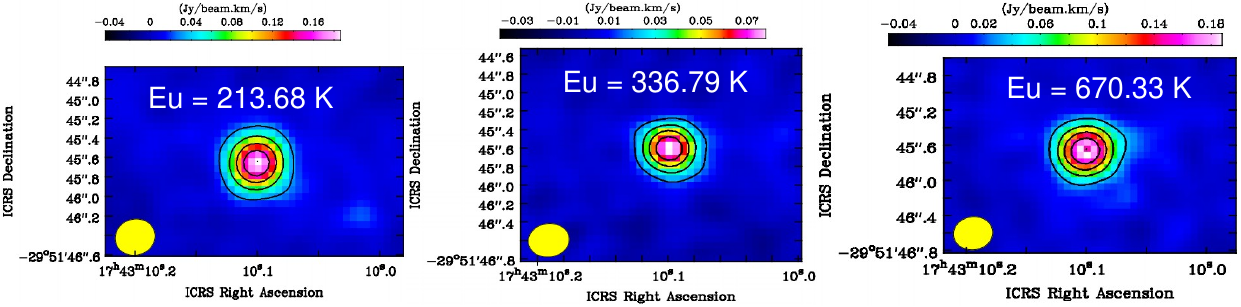}
	\includegraphics[width=1.0\textwidth]{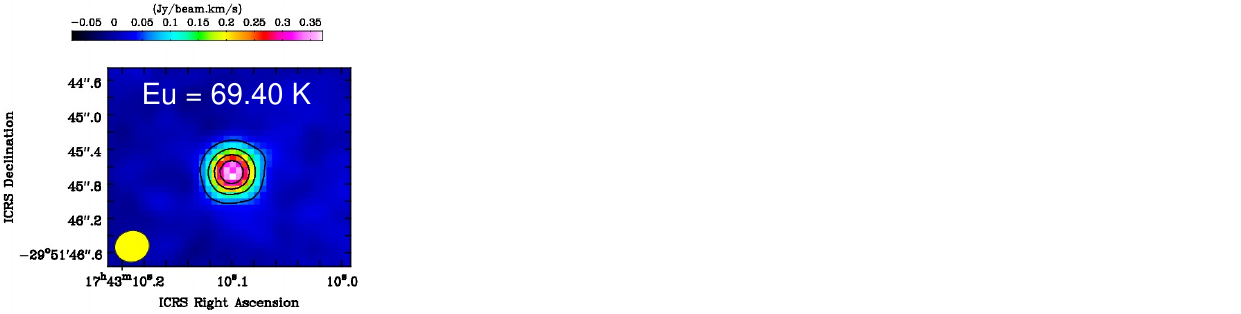}
	\caption{Integrated emission maps (moment zero) of \ce{CH2OHCHO} towards G358.93--0.03 MM1. The contour levels are at 20\%, 40\%, 60\%, and 80\% of the peak flux. Yellow circles represent the synthesized beams of the integrated emission maps.}
	\label{fig:emissionmap}
\end{figure*}

\subsubsection{Searching for \ce{CH2OHCHO} towards  G358.93--0.03 MM3}
After the successful detection of \ce{CH2OHCHO} in G358.93--0.03 MM1, we also search for emission lines of \ce{CH2OHCHO} towards G358.93--0.03 MM3, which yield no detection. The derived upper-limit column density of \ce{CH2OHCHO} towards this core is $\leq$(3.52$\pm$1.2)$\times$10$^{15}$ cm$^{-2}$. The upper limit of the fractional abundance is $\leq$(1.01$\pm$0.40)$\times$10$^{-8}$.

\subsection{Spatial distribution of \ce{CH2OHCHO} towards  G358.93--0.03 MM1}
We create the integrated emission maps (moment zero maps) of \ce{CH2OHCHO} towards  G358.93--0.03 MM1 using the CASA task {\tt IMMOMENTS}. In task {\tt IMMOMENTS}, we use channels corresponding to the velocity ranges, where the emission lines of \ce{CH2OHCHO} were detected. The integrated emission maps are shown in Figure~\ref{fig:emissionmap}. After the extraction, we apply the CASA task {\tt IMFIT} to fit the 2D Gaussian over the integrated emission maps of \ce{CH2OHCHO} to estimate the size of the emitting regions. The following equation is used

\begin{equation}        
\theta_{S}=\sqrt{\theta^2_{50}-\theta^2_{\text{beam}}}
\end{equation}
where $\theta_{\text{beam}}$ is the half-power width of the synthesized beam and $\theta_{50} = 2\sqrt{A/\pi}$ denotes the diameter of the circle whose area is surrounded by the $50\%$ line peak of \ce{CH2OHCHO} \citep{riv17}. The derived sizes of the emitting regions of \ce{CH2OHCHO} and velocity ranges at different frequencies are listed in Table~\ref{tab:emitting region}. The synthesized beam sizes of the integrated emission maps are 0.425$^{\prime\prime}\times$0.369$^{\prime\prime}$, 0.427$^{\prime\prime}\times$0.376$^{\prime\prime}$, 0.413$^{\prime\prime}\times$0.364$^{\prime\prime}$, and 0.410$^{\prime\prime}\times$0.358$^{\prime\prime}$, respectively. We observe that the estimated emitting region of \ce{CH2OHCHO} is comparable to or slightly greater than the synthesized beam sizes of the integrated emission maps. This indicates that the detected \ce{CH2OHCHO} transition lines are not spatially resolved or are only marginally resolved towards G358.93--0.03 MM1. Hence, we cannot draw any conclusions regarding the morphology of the spatial distributions of \ce{CH2OHCHO}. Higher spatial and angular resolution observations are required to understand the spatial distribution of \ce{CH2OHCHO} towards G358.93--0.03 MM1.
 
\section{Discussion}
\label{dis}
In this section, we compare the derived abundance of \ce{CH2OHCHO} in G358.93--0.03 MM1  with that of other hot cores and corinos. We also discuss the possible pathways for the formation of \ce{CH2OHCHO} in the context of hot molecular cores. Finally, we compare the observed abundance with those derived from chemical models.

\subsection{Comparison with other sources}
 We list the abundances of \ce{CH2OHCHO} towards IRAS 16293--2422 B, NGC 7129 FIRS 2, NGC 1333 IRAS2A, Sgr B2 (N), G31.41+0.31, and G10.47+0.0 taken from the literature in Table~\ref{tab:abundance}. We note that for NGC 1333 IRAS 2A, \cite{cou15} did not actually derive an abundance with respect to \ce{H2} but instead with respect to (\ce{CH2OH})$_{2}$ and \ce{CH3OCHO}. We, therefore, used their derived column density of \ce{CH2OHCHO} as well as the column density of \ce{H2} as derived by \citet{taq15} to infer an abundance with respect to \ce{H2}. Using the rotational diagram, \cite{cou15} derived the column density of \ce{CH2OHCHO} towards NGC 1333 IRAS2A as 2.4$\times$10$^{15}$ cm$^{-2}$ with a rotational temperature of 130 K. The column density of molecular hydrogen towards NGC 1333 IRAS2A is 5.0$\times$10$^{24}$ cm$^{-2}$ \citep{taq15}. To determine the fractional abundance of \ce{CH2OHCHO} with respect to \ce{H2} towards NGC 1333 IRAS2A, we use the column density of \ce{CH2OHCHO}, which is divided by the column density of \ce{H2}. We deduce a fractional abundance for \ce{CH2OHCHO} towards NGC 1333 IRAS2A with respect to \ce{H2} of 4.8$\times$10$^{-10}$. 

Our estimate for the abundance of \ce{CH2OHCHO} towards G358.93--0.03 MM1 ((4.90$\pm$2.92)$\times$10$^{-9}$) is quite similar to that of the hot molecular core G31.41+0.31 and the hot corino object IRAS 16293--2422 B while approximately one order of magnitude higher than that of G10.47+0.03, Sgr B2 (N), NGC 1333 IRAS 2A, and NGC 7129 FIRS 2. The similarity among G358.93--0.03 MM1, G31.41+0.31, and IRAS 16293--2422 B may indicate that the formation route(s) of \ce{CH2OHCHO} may be similar in all three sources.
 	
 \subsection{Possible formation mechanisms of \ce{CH2OHCHO} towards  hot molecular cores and hot corinos}
To date, only a few efficient formation pathways of \ce{CH2OHCHO} have been proposed on grain surfaces in hot molecular cores and hot corinos \citep{gar06, gar08, gar13, cou18, riv19, min20}. In the high temperature ($\geq 100$ K) regime, the radicals gain sufficient energy to diffuse across the surface and react to create complex organic molecules \citep{min20}. Initially, two main formation pathways were proposed for the formation of \ce{CH2OHCHO}:\\\\
HCO + \ce{CH2OH}$\rightarrow$\ce{CH2OHCHO}~~~~~~~~~~~~(1)\\\\
and\\\\
\ce{CH3OH} + HCO$\rightarrow$\ce{CH2OHCHO}+H~~~~~~(2)\\\\
In reaction 1, radical HCO and radical \ce{CH2OH} react with each other on the grain surfaces to form \ce{CH2OHCHO} \citep{gar08,gar13, cou18, riv19}. This reaction appears to be responsible for the production of \ce{CH2OHCHO} towards IRAS 16293--2422 B and G31.41+0.31 \citep{jo12, riv19, min20}. In reaction 2, the radicals HCO and \ce{CH3OH} react with each other on the grain surfaces to produce \ce{CH2OHCHO}, but this reaction has not yet been tested in a laboratory \citep{min20}. Initially, \citet{fed15} and \citet{chu16} experimentally studied the possible formation pathways of \ce{CH2OHCHO}  on dust grains at low temperatures ($\sim$10 K).
The experimental results of \citet{fed15} and \citet{chu16} were later confirmed by \citet{sim20} by using microscopic kinetic Monte Carlo simulations based on ice chemistry.

\begin{table}
	%	\begin{minipage}[t]{\columnwidth}
	\centering
	%	\scriptsize 
	\caption{Estimated emitting regions of \ce{CH2OHCHO} towards the G358.93--0.03 MM1.}
	\begin{adjustbox}{width=0.45\textwidth}
		\begin{tabular}{ccccccccccccccccc}
			\hline 
			Observed frequency& $E_{u}$& emitting region&Velocity ranges\\
			(GHz) & (K)&($^{\prime\prime}$)&(km s$^{-1}$) \\
			\hline
			~290.589$^{*}$&182.37&0.410&--14.29 to --19.61 \\
			
			291.094&207.92&0.412&--13.39 to --18.68\\
			
			291.784&211.41&0.413&--14.38 to --20.57\\

			292.536&213.68&0.414&--13.35 to --19.31 \\

			292.737&336.79&0.415&--14.40 to --19.22 \\
			
			305.040&670.33&0.411&--13.34 to --18.44 \\
			
			~305.488$^{*}$&~69.40&0.410&--14.14 to --18.79 \\
			
			\hline
		\end{tabular}	
	\end{adjustbox}
	{{*}}--There are two transitions that have close frequencies ($ \leq 100$ kHz), and only the frequency of the first transition is shown.\\
	\label{tab:emitting region}
	%	\end{minipage}[t]{\columnwidth}
\end{table}

\subsection{Chemical modelling of \ce{CH2OHCHO} in  hot molecular cores}
 
To understand the formation mechanisms and abundance of \ce{CH2OHCHO} in hot molecular cores, \cite{cou18} computed a two-phase warm-up chemical model using the gas grain chemistry code {\tt UCLCHEM} \citep{hol17}. They assumed a free-fall collapse of a cloud (Phase I), followed by a warm-up phase (Phase II). In the first phase (Phase I), the gas density increased from $n_{H}$ = 300 cm$^{-3}$ to 10$^{7}$ cm$^{-3}$, and they assumed a constant dust temperature of 10 K. In the second phase (phase II), the gas density remained constant at 10$^{7}$ cm$^{-3}$, whereas the dust temperature increased with time from 10 to 300 K. This phase was known as the warm-up phase. In the chemical network used by \citet{cou18}, the recombination of the radicals HCO and \ce{CH2OH} (Reaction 1) dominates the production of \ce{CH2OHCHO} on the grains. Reaction 1 is the most likely pathway because \cite{but15} tested this reaction in the laboratory and confirmed that the reaction produced \ce{CH2OHCHO}. In the warm-up phase, \citet{cou18} showed that the abundance of \ce{CH2OHCHO} varied from $\sim$10$^{-9}$ to 10$^{-8}$ (see Figure 3 in \citet{cou18}). \citet{cou18} did not include reaction 2 (recombination of the radical HCO and \ce{CH3OH}) as earlier work by \cite{wo12} showed that by including reaction 2, the estimated model abundance of \ce{CH2OHCHO} was as high as $\sim$10$^{-5}$ \citep{wo12}. This modelled abundance of \ce{CH2OHCHO} by reaction 2 does not match any of the observed abundances in the sample of objects considered here.
	
\subsection{Comparison between observed and chemically modelled abundance of \ce{CH2OHCHO}}
In order to understand the formation pathways of \ce{CH2OHCHO} towards G358.93--0.03 MM1, we compare our estimated abundance with the modelled ones from \cite{cou18}. This comparison is physically reasonable because the dust temperature of this source is 150 K, which is a typical hot core temperature, and the number density (n$_{H}$) of this source is $\sim$2$\times$10$^{7}$ cm$^{-3}$ \citep{chen20, ste21}. Hence, the two-phase warm-up chemical model based on the timescales in \cite{cou18} is appropriate for explaining the chemical evolution of \ce{CH2OHCHO} towards G358.93--0.03 MM1. \cite{cou18} showed that the abundances of \ce{CH2OHCHO} varies between $\sim$10$^{-9}$ and 10$^{-8}$. We find that our estimated abundance towards G358.93--0.03 MM1 is (4.90$\pm$2.92)$\times$10$^{-9}$, which is in good agreement with the theoretical results in \cite{cou18}. This comparison indicates that the simplest sugar-like molecule, \ce{CH2OHCHO}, may form on the grain surface via the reaction between radical HCO and radical \ce{CH2OH} (Reaction 1) towards G358.93--0.03 MM1. Of course, the modelled abundance of \ce{CH2OHCHO} is also similar to the observed abundance of \ce{CH2OHCHO} towards the hot molecular core G31.41+0.31 and the hot corino object IRAS 16293--2422 B, indicating that reaction 1 may be the most likely pathway for the production of \ce{CH2OHCHO} towards these two objects too. Radical HCO and radical \ce{CH2OH} may be created in the ISM by the hydrogenation of CO (CO + H$\rightarrow$HCO$^{\bullet}$+H$\rightarrow$\ce{H2CO}$\rightarrow$\ce{$^{\bullet}$CH2OH}) \citep{ha13}. After hydrogenation, radical \ce{CH2OH} is converted into \ce{CH3OH} (\ce{$^{\bullet}$CH2OH} + H$\rightarrow$\ce{CH3OH}) \citep{ha13}. Our conclusion agrees with the recent work of \citet{min20}, who also found that reaction 1 is the most efficient pathway for the formation of \ce{CH2OHCHO} towards the hot core G31.41+0.31, as well as other hot molecular cores.

\begin{table}
	%	\begin{minipage}[t]{\columnwidth}
	\centering
	%	\scriptsize 
	\caption{Abundance of \ce{CH2OHCHO} in different objects.}
	\begin{adjustbox}{width=0.45\textwidth}
		\begin{tabular}{ccccccccccccccccc}
			\hline 
Source&X(\ce{CH2OHCHO})& Refreances\\
Name  &			       &           \\
			\hline
G358.93--0.03 MM1&(4.90$\pm$2.92)$\times$10$^{-9}$ &This paper \\
IRAS 16293--2422 B&5.8$\times$10$^{-9}$ &\citep{jo12} \\
NGC 7129 FIRS 2  &5.0$\times$10$^{-10}$ &\citep{flu14} \\
NGC 1333 IRAS2A  &4.8$\times$10$^{-10}$ &See section~4.1\\
 Sgr B2 (N)      &1.6$\times$10$^{-10}$ &\citep{xu19} \\
 G31.41+0.31     &(5.0$\pm$1.4)$\times$10$^{-9}$ &\citep{min20} \\
 G10.47+0.03	& 9.6$\times$10$^{-10}$ & \citep{mondal21}\\
			
			\hline
		\end{tabular}	
	\end{adjustbox}
	\label{tab:abundance}
	%	\end{minipage}[t]{\columnwidth}
\end{table}

\section{Conclusions}
\label{conclu} 
We present the first detection of \ce{CH2OHCHO} using ALMA in the hot molecular core G358.93--0.03 MM1. We identify a total of seventy-five transitions of \ce{CH2OHCHO}, where the upper-level energies vary between 63.86 K and 1362.19 K. The derived abundance of \ce{CH3OCHO} is (4.90$\pm$2.92)$\times$10$^{-9}$. We compare our estimated abundance with that of other hot molecular cores and hot corinos and note that the abundance of \ce{CH2OHCHO} towards G358.93--0.03 MM1 is quite similar to that found towards another hot molecular core, G31.41+0.31, and the hot corino, IRAS 16293--2422 B \citep{jo12, min20}. We discuss the possible formation mechanisms of \ce{CH2OHCHO} in hot molecular cores. We compare our estimated abundance of \ce{CH2OHCHO} with the theoretical abundance from the chemical model presented in \citet{cou18} and find that they are similar. We conclude that  \ce{CH2OHCHO}  is most likely formed via the reaction of radical HCO and radical \ce{CH2OH} on the grain surfaces in G358.93--0.03 MM1 and other hot molecular cores.
	
The identification of abundant \ce{CH2OHCHO} in G358.93--0.03 MM1 suggests that grain surface chemistry is also efficient for the formation of other complex organic molecules in this hot molecular core, including isomers of \ce{CH2OHCHO}, \ce{CH3OCHO} and \ce{CH3COOH}. Indeed, the highly chemically rich spectra and detection of \ce{CH2OHCHO} towards G358.93--0.03 MM1 make this object another ideal hot core to search for and study other complex organic molecules in star-forming regions. A spectral line study combined with a radiative transfer as well as a two-phase warm-up chemical model is required to understand the prebiotic chemistry of G358.93--0.03 MM1, which will be carried out in our follow-up study.

\section*{ACKNOWLEDGEMENTS}
We thank the anonymous referee for her/his constructive comments that helped improve the manuscript. A.M. acknowledges the Swami Vivekananda Merit-cum-Means Scholarship (SVMCM) for financial support for this research. SV acknowledges support
from the European Research Council (ERC) under the European Union's Horizon 2020 research and innovation program MOPPEX
833460. This paper makes use of the following ALMA data: ADS /JAO.ALMA\#2019.1.00768.S. ALMA is a partnership of ESO (representing its member states), NSF (USA), and NINS (Japan), together with NRC (Canada), MOST and ASIAA (Taiwan), and KASI (Republic of Korea), in co-operation with the Republic of Chile. The Joint ALMA Observatory is operated by ESO, AUI/NRAO, and NAOJ.

\section*{DATA AVAILABILITY}
The plots within this paper and other findings of this study are available from the corresponding author on reasonable request. The data used in this paper are available in the ALMA Science Archive (\url{https://almascience.nrao.edu/asax/}), under project code 2019.1.00768.S.

\bibliographystyle{aasjournal}
%\bibliography{./literature.bib,added.bib} % if your bibtex file is called example.bib

% Don't change these lines
\bsp	% typesetting comment
\label{lastpage}
\end{document}